\documentstyle[12pt]{article}
\def\ba{\begin{eqnarray}}
\def\ea{\end{eqnarray}}

\def\lb{\label}
\def\be{\begin{equation}}
\def\ee{\end{equation}}

\begin{document}

\begin{center}
{\Large \bf Operator approach to analytical evaluation of Feynman
diagrams}\footnote{This work was supported by grant RFBR 05-01-01086-a.}
\end{center}


\begin{center}
\large{A.P.\,Isaev}
\end{center}

\begin{center}
Bogoliubov Laboratory of Theoretical Physics, \\
Joint Institute for Nuclear Research, \\
Dubna, Moscow region 141980, Russia \\
E-mail: isaevap@theor.jinr.ru
\end{center}


{\bf Abstract.}
The operator approach to analytical evaluation of
multi-loop Feynman diagrams is proposed.
We show that the known analytical methods of evaluation of
massless Feynman integrals,
such as the integration by parts method and
the method of "uniqueness" (which is based on the star-triangle
relation), can be drastically simplified by
using this operator approach.
To demonstrate the
advantages of the operator method of analytical evaluation of
multi-loop Feynman diagrams, we calculate ladder diagrams for the massless $\phi^3$
theory (analytical results for these diagrams
are expressed in terms of multiple polylogarithms).
It is shown how operator formalism can be applied to calculation
of certain massive Feynman diagrams and investigation of
Lipatov integrable chain model.

\section{Introduction}
\setcounter{equation}0

Elaboration of methods of analytical evaluation of multiple integrals
visualized as multi-loop
Feynman diagrams is motivated from the point of view
of physical applications and the mathematical point of view.

{}From the physical point of view such investigations are important, since
in calculations of the physical characteristics in quantum field
theories the number of Feynman diagrams,
in higher order of the perturbation theory, grows so quickly \cite{BogShi} that
numerical calculations are not sufficient to obtain the desirable precision
for the corresponding sums of perturbative integrals.

{}From the point of view of mathematics and mathematical physics such
investigations are interesting since they reveal some structures specific
for the theory of quantum integrable systems
\cite{Zam}, \cite{Lip}, \cite{Kucha}
(see also \cite{BeSt} and references therein). Besides, as a rule, analytical results
are expressed in terms of multiple zeta-values and polylogarithms -- special
functions which are extremely interesting and important objects of
investigations in modern mathematics
and mathematical physics (see, e.g., \cite{Zag}, \cite{Gonch}, \cite{Kreim}).

{}Finally, as we will see in this paper,
the analytical results in calculations of Feynman diagrams (FDs)
give explicit expressions for Green functions of certain special
quantum mechanical models. And vice versa, the expansion of such
Green functions over coupling constants gives explicit expressions
for multi-loop FDs. This connection of perturbative Feynman integrals and Green functions for
integrable quantum mechanical models is one of the substantial
implications of the proposed algebraic approach.

The main idea of our algebraic method for evaluation
of FDs is that we replace manipulations with multiple integrals by
manipulations with the corresponding algebraic expressions.
In other words, identical transformations of multiple perturbative integrals
are substituted with transformations of generators of special infinite dimensional
algebra. This drastically simplifies all calculations.
We stress that here we mostly consider the massless case of FDs (see, however, Subsection 4.3).

\section{Feynman diagrams in configuration space}
\setcounter{equation}0

The Feynman diagrams, which will be considered in this paper, are graphs
with vertices connected by edges (propagators). To each edge we assign a
complex number (the index of the propagator).
With each vertex we associate the point in the $D$-dimensional space ${\bf R}^D$
while the edges of the graph (with index $\alpha$) are associated with the
propagator of massless particle

\unitlength=1cm
\begin{picture}(25,1.5)

\put(1,0.5){\line(1,0){1.8}}
\put(1.7,0.7){$\alpha$}

\put(0.5,0.45){$x$}
\put(3.2,0.45){$y$}
\put(5,0.45){$=$}
\put(6,0.45){$1 /(x-y)^{2\alpha}$}
\end{picture}

\noindent
where $\; x,y \in {\bf R}^D$, $(x-y)^{2\alpha} := (\sum_{i=1}^D  (x_i-y_i) \, (x_i-y_i))^{\alpha}$,
and $\; \alpha \in {\bf C}$. Moreover, we consider the graphs with two types
of vertices: boldface vertices $\bullet$ denote that the corresponding
points are integrated over ${\bf R}^D$. These FDs are called
{\it FDs in the configuration space}. Note that more commonly known
{\it the momentum space FDs} are \underline{dual graphs} with respect to the graphs
which correspond to the FD in the configuration space.
That is, loops in the momentum space FD are replaced by boldface vertices in FD
in the configuration space, and vice versa, vertices in the momentum space FD are replaced by loops in
FD in the configuration space.

Consider examples of FDs in the configuration space and present
the corresponding multiple integrals.

\vspace{0.2cm}

{\bf 1. Graph with 5 vertices and 5 edges (3-point function)}:

\unitlength=5mm
\begin{picture}(25,4)

\put(0,1){\line(1,0){6}}
\put(2.7,1.2){\tiny $\alpha_3$}
\put(3,3.3){$0$}

\put(2,1){\line(1,2){1}}
\put(1.9,0.85){$\bullet$}
\put(1.9,0.45){$z$}

\put(0.8,1.2){\tiny $\alpha_1$}
\put(1.7,2){\tiny $\alpha_2$}

\put(4,1){\line(-1,2){1}}
\put(3.9,0.85){$\bullet$}
\put(3.9,0.45){$u$}

\put(4.8,1.2){\tiny $\alpha_5$}
\put(3.8,2){\tiny $\alpha_4$}

\put(-0.1,0.5){$x$}
\put(5.9,0.5){$y$}
\put(8,1.5){$=$}
\put(9,1.5){$ \int \frac{d^D z \; d^D u}{(z-y)^{2\alpha_1} \,
z^{2\alpha_2} \, y^{2\alpha_3} \, u^{2\alpha_4}\, (u-y)^{2\alpha_5}}$ }

\end{picture}

{\bf 2. "Star" graph}:

\vspace{0.5cm}

\unitlength=4.5mm
\begin{picture}(20,4.9)

\put(5,1){\line(2,1){2}}
\put(7,2){\line(2,-1){2}}
\put(7,1.9){\line(0,1){2}}
\put(6.85,1.8){$\bullet$}

\put(8.1,1.6){\tiny $\alpha_3$}
\put(5.2,1.6){\tiny $\alpha_1$}
\put(6.1,3){\tiny  $\alpha_2$}
\put(6.8,1.3){$x$}
\put(6.7,4.3){$x_2$}
\put(4.8,0.5){$x_1$}
\put(9,0.5){$x_3$}

\put(11,2){$=$}

\put(13,2){$\int \frac{d^D x}{(x-x_1)^{2\alpha_1} \,
(x-x_2)^{2\alpha_2} \, (x-x_3)^{2\alpha_3} }$}

\end{picture}

{\bf 3. Propagator type graph (2-point function):}

\vspace{0.2cm}

\unitlength=5mm
\begin{picture}(25,4)

\put(1,2){\line(1,0){1.8}}
\put(3.2,2){\line(1,0){1.8}}
\put(2.0,1.7){\tiny  $\alpha_6$}
\put(1,2){\line(2,1){2}}
\put(1,2){\line(2,-1){2}}
\put(1.6,2.85){\tiny  $\alpha_4$}
\put(1.7,1){\tiny  $\alpha_2$}
\put(3.05,2.3){\tiny  $\alpha_3$}

\put(0.5,1.6){$0$}

\put(3,1){\line(0,1){2}}
\put(2.85,0.85){$\bullet$}
\put(2.85,2.85){$\bullet$}
\put(4.85,1.85){$\bullet$}

\put(3,3){\line(2,-1){2}}
\put(4,2.5){\tiny  $\alpha_5$}
\put(3,1){\line(2,1){2}}
\put(4,1){\tiny  $\alpha_1$}

\put(5.35,3.2){\line(-1,-3){0.4}}
\put(5.2,2.4){\tiny  $\alpha_7$}
\put(5.3,3.1){\line(-1,0){2.2}}
\put(4,3.3){\tiny $\alpha_8$}
\put(6,3.3){\tiny  $\alpha_9$}
\put(5.2,2.9){$\bullet$}
\put(5.3,3.1){\line(1,0){1.7}}

\put(5.3,1.6){$y$}
\put(5.1,3.3){$w$}
\put(7.2,3){$x$}
\put(2.7,0.4){$z$}
\put(2.7,3.3){$u$}

\put(7.8,2){$=$}
\put(8.8,2){$\int \frac{d^D z \; d^D u \; d^D y \; d^D w}{(x-z)^{2\alpha_1} \,
z^{2\alpha_2} \, (z-u)^{2\alpha_3} \, u^{2\alpha_4}\, (u-y)^{2\alpha_5} \, y^{2\alpha_6}
\dots (w-x)^{2\alpha_9}}$}

\end{picture}

\vspace{-0.1cm}

The problem (which we need to solve when analytically calculate multiple
integrals corresponding to FDs) consists in searching
for special transformations of graphs (FDs) such that the number of
boldface vertices (the number of integrations) decreases at each step.
In the next section, we discuss these special transformations and describe
the corresponding operator formalism, which gives us a possibility to
represent these transformations using a more compact algebraic language.

\section{Operator formalism}
\setcounter{equation}0

\subsection{Algebraic manipulations with perturbative integrals}

Consider the $D$-dimensional Euclidean space ${\bf R}^D$ with the coordinates $x_i$,
$(i=1,2, \dots, D)$. We use the notation: \fbox{$x^{2\alpha} = (\sum_{i=1}^D x_i^2 )^{\alpha}$}.
Let the operators of coordinates $\hat{q}_i = \hat{q}_i^\dagger$
and momenta $\hat{p}_i = \hat{p}_i^\dagger$ be generators of the Heisenberg algebra
$$
[\hat{q}_k , \, \hat{p}_j ] = {\rm i} \, \delta_{kj} \; .
$$
Introduce the vectors $|x \rangle \equiv |\{ x_i \}\rangle$, $|k\rangle\equiv |\{ k_i \}\rangle$
that are eigenstates of the operators of coordinates and momenta, respectively:
$\hat{q}_i |x\rangle  = x_i \, |x\rangle$, $\hat{p}_i |k\rangle  = k_i \, |k\rangle$.
We normalize these states as follows
$$
\langle x|k \rangle = \frac{1}{(2 \pi)^{D/2}} \, \exp ({\rm i} \, k_j \, x_j ) \; , \;\;\;
 \int d^D k \, |k\rangle \, \langle k| = \hat{1} = \int d^D x \, |x\rangle \, \langle x| \; .
$$
Consider the heat-kernels ("matrix representations") of
the operators $\hat{p}^{-2\beta}$:
\begin{equation}
\label{matrep}
\langle x| \frac{1}{\hat{p}^{2 \beta}} |y\rangle = a(\beta) \,
\frac{1}{(x-y)^{2 \beta'}}
 \; , \;\;\;\; \left( a(\beta) =
\frac{\Gamma(\beta') }{ \pi^{D/2} \, 2^{2\beta} \, \Gamma(\beta)} \right) .
\end{equation}
where \fbox{$\beta' = D/2 - \beta$} and $\Gamma(\beta)$ is the Euler gamma-function. Formula
(\ref{matrep}) relates the propagators for massless particles and pseudo-differential operators
$\hat{p}^{-2\beta}$. For the operators $\hat{q}^{2\alpha}$
the "matrix representations" have the form:
\begin{equation}
\label{matrep2}
\langle x| \hat{q}^{2 \alpha} |y\rangle = x^{2\alpha} \, \delta^D(x-y) \; .
\ee

\vspace{0.2cm}

Below we consider three
{\bf (a,b,c)} \underline{algebraic relations} which are operator analogs
of relations used for the analytical evaluation of multi-loop
perturbative integrals for FDs. Recall that these relations give us a possibility
to reconstruct FD in such a way that the number of integrations
(the number of boldface vertices in the graph) decreases to zero
(this will indicate that a given FD is calculated analytically).

\underline{\bf a. Group relation}. Consider a convolution product of two propagators:
\be
\lb{convol}
\int  \frac{d^D z \,}{(x-z)^{2 \alpha} \, (z-y)^{2 \beta}} =
\frac{G(\alpha' ,\beta')}{(x-y)^{2(\alpha + \beta - D/2)}}
\; , \;\;\; \left( G(\alpha ,\beta) = \frac{a ( \alpha + \beta )}{a(\alpha) \, a(\beta)} \right) \, .
\ee
We graphically represent this convolution product as

\vspace{0.2cm}

\unitlength=8mm
\begin{picture}(25,1.8)

\put(1,0.7){\line(1,0){3.3}}
\put(2.5,0.6){$\bullet$}
\put(1.7,0.9){\small $\alpha$}
\put(3.5,0.9){\small $\beta$}

\put(0.5,0.65){$x$}
\put(4.7,0.65){$y$}
\put(2.5,0.3){$z$}

\put(6,0.65){$=$}

\put(7,0.65){$G(\alpha' ,\beta')\; \cdot$}

\put(11,0.7){\line(1,0){2.5}}
\put(11.5,0.9){\scriptsize $\alpha \! + \! \beta \! - \! \frac{D}{2}$}

\put(10.5,0.65){$x$}
\put(13.7,0.65){$y$}


\end{picture}

\noindent
Thus, relation (\ref{convol}) describes such a reconstruction of the graph in which
the number of integrations (boldface vertices) decreases by one.
This relation is the "matrix representation" of the operator identity
(group relation)
\be
\lb{convol1}
\hat{p}^{-2 \alpha'} \, \hat{p}^{-2 \beta'} = \hat{p}^{-2 (\alpha' + \beta')}.
\ee
Indeed, by using
(\ref{matrep}) and (\ref{matrep2}) we easily demonstrate that the "matrix" analog of (\ref{convol1})
$$
\int \, d^D  z \; \langle x | \hat{p}^{-2 \alpha'} \, | z \rangle \, \langle z |
 \, \hat{p}^{-2 \beta'} | y \rangle =
\langle x | \hat{p}^{-2 (\alpha' + \beta')} | y \rangle
$$
coincides with relation (\ref{convol}).
Note that in the operator relation
(\ref{convol1}) the tedious coefficient $G(\alpha',\beta')$ is vanished.

\vspace{0.25cm}

\underline{\bf b. Star-triangle relation}. This relation is in the basis of the so-called
"method of uniqueness" \cite{Kaz}
(see also \cite{Vas}) -- an efficient method of analytical evaluation of FD.
In fact, this relation is a special case of the Yang-Baxter equation \cite{Zam}, \cite{Is}.
The star-triangle relation (STR) has the form
\be
\lb{zvtr}
\int  \frac{d^D z \,}{(x-z)^{2 \alpha'} \, z^{2 (\alpha + \beta)} \,
(z-y)^{2 \beta'} } =
\frac{G(\alpha ,\beta)}{(x)^{2 \beta} \, (x-y)^{2(\frac{D}{2}-\alpha-\beta)} \,
(y)^{2 \alpha} }
\; ,
\ee
and was initially used  in the framework of investigations of multi-dimensional
conformal field theories \cite{FP}. The identity (\ref{zvtr}) can be graphically
represented as

\unitlength=4.5mm
\begin{picture}(20,4.9)

\put(2,1){\line(2,1){2}}
\put(4,2){\line(2,-1){2}}
\put(4,1.9){\line(0,1){2}}
\put(3.9,1.8){$\bullet$}

\put(5.1,1.6){\tiny $\beta'$}
\put(2.3,1.6){\tiny $\alpha'$}
\put(2.8,3){\tiny $\alpha \! +\! \beta$}
\put(3.8,1.3){$z$}
\put(3.7,4.3){$0$}
\put(1.8,0.5){$x$}
\put(6,0.5){$y$}

\put(6.9,2.2){$=  G(\alpha ,\beta)\; \cdot$}

\put(12,1){\line(1,0){3}}
\put(13.2,4.3){$0$}
\put(11.5,0.6){$x$}
\put(15.4,0.6){$y$}

\put(12,1){\line(1,2){1.5}}
\put(12.2,2.5){\tiny $\beta$}
\put(12.5,0.5){\tiny $(\alpha \! +\! \beta)'$}
\put(14.4,2.5){\tiny $\alpha$}
\put(15,1){\line(-1,2){1.5}}


\end{picture}

\noindent
Thus, STR (\ref{zvtr}) describes such a reconstruction of the graph for which
the number of integrations (boldface vertices) decreases by one.
The operator version of this relation was proposed in \cite{Isa03}
and is written in the form
\be
\lb{zvtr2}
\hat{p}^{-2 \alpha} \hat{q}^{-2(\alpha+\beta)} \hat{p}^{-2 \beta} =
\hat{q}^{-2 \beta} \hat{p}^{-2(\alpha+\beta)} \hat{q}^{-2 \alpha} \;\;\; (\forall \alpha,\beta) \; ,
\ee
Again we note the absence of the coefficient $G(\alpha,\beta)$.
To demonstrate the equivalence of (\ref{zvtr}) and (\ref{zvtr2}) we
act on (\ref{zvtr2}) by the vectors $\langle x|$ and
$| y \rangle$ from the left and right, respectively,
and use the representations (\ref{matrep}), (\ref{matrep2}).
Identity (\ref{zvtr2}) can be written in the form
of the Yang-Baxter equation:
$$
R_{12}(\alpha) \, R_{23}(\alpha + \beta) \, R_{12}(\beta) =
R_{23}(\beta) \, R_{12}(\alpha + \beta) \, R_{23}(\alpha) \; ,
$$
where $R_{ab}(\alpha)= (\hat{q}^{(a)} - \hat{p}^{(b)})^{2\alpha}$ and
$[\hat{q}^{(a)}_i ,  \, \hat{p}^{(b)}_j ] = {\rm i} \, \delta_{ij} \, \delta^{ab}$,
$(i,j = 1,\dots,D)$.
Now a few remarks about STR (\ref{zvtr2}) are in order. \\
{\bf Remark 1.} The algebraic version of STR is equivalent to
the commutativity for the infinite set of operators
$H(\alpha) = \hat{p}^{2\alpha} \hat{q}^{2\alpha}$:
$$
H(\alpha) \, H^{-1}(-\beta) =  H^{-1}(-\beta) \, H(\alpha) \; \Rightarrow \;
 \hat{p}^{2\alpha} \hat{q}^{2(\alpha + \beta)}  \, \hat{p}^{2\beta}  =
\hat{q}^{2\beta} \, \hat{p}^{2(\alpha+\beta)} \, \hat{q}^{2\alpha} \; .
$$
{\bf Remark 2.} Here we present the algebraic proof of STR (\ref{zvtr2}). Introduce
an inversion operator ${\cal R}_\Delta$ which obeys the conditions
\be
\lb{svR}
{\cal R}_\Delta^2=1  \; , \;\;\;
{\cal R}_\Delta \, \hat{q}_i \, {\cal R}_\Delta = \hat{q}_i \, / \, \hat{q}^2 \; , \;\;\;
\langle x_i | \, {\cal R}_\Delta = \langle {x_i \over x^2} | x^{2\frac{(\Delta-D)}{2}}\; ,
\ee
\be
\lb{svR1}
\begin{array}{c}
{\cal R}_\Delta^\dagger = {\cal R}_\Delta \hat{q}^{2\Delta} \, ,  \;\;
K_j^{(\Delta)} := {\cal R}_\Delta \, \hat{p}_j \, {\cal R}_\Delta =
 \hat{q}^2 \, \hat{p}_j - 2 \, \hat{q}_j \, (\hat{q} \, \hat{p}) +
{\rm i} (D-\Delta) \hat{q}_j  \, ,   \\ [0.3cm]
{\cal R}_\Delta \, \hat{p}^{2 \beta} \, {\cal R}_\Delta =
\hat{q}^{2 (\beta + { \Delta \over 2} )} \, \hat{p}^{2 \beta}  \,
\hat{q}^{2 (\beta - {\Delta \over 2} )} \, .
\end{array}
\ee
We note that
${\cal R}_\Delta = {\cal R} \hat{q}^{-2\frac{\Delta}{2}}$, where ${\cal R} \equiv {\cal R}_0$.
Using (\ref{svR}), (\ref{svR1}) the algebraic version of STR is proved as follows:
$$
\begin{array}{cc}
{\cal R} \, \hat{p}^{2\alpha} \,
\hat{p}^{2\beta} \, {\cal R}  & = {\cal R} \, \hat{p}^{2(\alpha +\beta)} \, {\cal R}
\;\; \Rightarrow \;\;
\hat{p}^{2\alpha} \hat{q}^{2(\alpha +\beta)}  \, \hat{p}^{2\beta}  =
 \hat{q}^{2\beta} \, \hat{p}^{2(\alpha +\beta)} \hat{q}^{2\alpha} \; . \\ [-0.2cm]
\!\!\!\! \Uparrow & \\
{\cal R}^2 & \\ [-0.4cm]
\end{array}
$$
{\bf Remark 3.}
From operator identity (\ref{zvtr2}) one can immediately obtain new STR
\be
\lb{myybe}
\begin{array}{c}
(\hat{p}^{2 \alpha} + a \, \hat{q}^{-2 \alpha})( \hat{q}^{2 (\alpha +\beta)} +
b \, \hat{p}^{-2 (\alpha +\beta)})
( \hat{p}^{2 \beta} + c \, \hat{q}^{-2 \beta}) =  \\ [0.2cm]
=( \hat{q}^{2 \beta} + c \, \hat{p}^{-2 \beta})( \hat{p}^{2 (\alpha +\beta)} +
b \, \hat{q}^{-2 (\alpha +\beta)})
(\hat{q}^{2 \alpha} + a \, \hat{p}^{-2 \alpha}) \; ,
\end{array}
\ee
where $a,b,c$ are arbitrary constants. \\
{\bf Remark 4.} One can introduce one more "local" STR \cite{Kash}
which is related to the $\alpha$-representation of perturbative Feynman integrals
$$
W(\hat{q}^2 |\alpha_1) \, W(\hat{p}^2 |\, \frac{1}{\alpha_2}) \,
W(\hat{q}^2 |\alpha_3) =
W(\hat{p}^2 |\, \frac{1}{\beta_3}) \, W(\hat{q}^2 |\beta_2) \, W(\hat{p}^2 |\, \frac{1}{\beta_1}) \; ,
$$
where
$W(x^2 | \alpha) = \exp \left( -x^2/(2\alpha) \right)$,
and the parameters $\alpha_i$ and $\beta_i$ are related by the identity
$\alpha_i = \frac{\beta_1 \beta_2 + \beta_1 \beta_3 + \beta_2 \beta_3}{\beta_i}$ which is known
as star-triangle transformation for resistances in electric networks.

\vspace{0.3cm}

\underline{\bf c. Integration by parts rule} \cite{ChTk1}.

First, we present the graphical version of this rule

\unitlength=4.5mm
\begin{picture}(20,5.5)

\put(0,1){\line(2,1){2}}
\put(2,2){\line(2,-1){2}}
\put(2,1.9){\line(0,1){2}}
\put(1.85,1.8){$\bullet$}

\put(3.1,1.6){\tiny $\alpha_3$}
\put(0.3,1.6){\tiny $\alpha_1$}
\put(1.2,3){\tiny $\alpha_2$}
\put(1.7,4.3){$0$}
\put(-0.2,0.5){$x$}
\put(4,0.5){$y$}

\put(4.5,2.2){$=  \frac{1}{(D- 2\alpha_2 - \alpha_1 - \alpha_3)}\;\;
\left\{ \alpha_1  \left( \right.\right.$}

\put(14,1){\line(2,1){2}}
\put(16,2){\line(2,-1){2}}
\put(16,1.9){\line(0,1){2}}
\put(15.85,1.8){$\bullet$}

\put(17.1,1.6){\tiny $\alpha_3$}
\put(13.8,1.6){\tiny $\alpha_1\! +\! 1$}
\put(14.5,3){\tiny $\alpha_2 \!- \! 1$}
\put(15.7,4.3){$0$}
\put(13.8,0.5){$x$}
\put(18,0.5){$y$}

\put(19,2.2){$-$}

\put(20,1){\line(2,1){2}}
\put(22,2){\line(2,-1){2}}
\put(22,1.9){\line(0,1){2.1}}
\put(21.85,1.8){$\bullet$}

\put(22,4){\line(-2,-3){2}}

\put(20.3,2.7){\tiny $-1$}

\put(23.1,1.6){\tiny $\alpha_3$}
\put(20.7,1.1){\tiny $\alpha_1\! +\! 1$}
\put(22.1,3){\tiny $\alpha_2$}
\put(21.7,4.3){$0$}
\put(19.8,0.5){$x$}
\put(24,0.5){$y$}

\put(25,2.2){$) \; +$}

\end{picture}

\unitlength=4.5mm
\begin{picture}(20,4.9)

\put(4.5,2.2){$+ \alpha_3 \; ($}

\put(7,1){\line(2,1){2}}
\put(9,2){\line(2,-1){2}}
\put(9,1.9){\line(0,1){2}}
\put(8.85,1.8){$\bullet$}

\put(10.1,1.6){\tiny $\alpha_3\! +\! 1$}
\put(7.2,1.6){\tiny $\alpha_1$}
\put(7.5,3){\tiny $\alpha_2 \!- \! 1$}
\put(8.7,4.3){$0$}
\put(6.8,0.5){$x$}
\put(11,0.5){$y$}

\put(12,2.2){$-$}

\put(13,1){\line(2,1){2}}
\put(15,2){\line(2,-1){2}}
\put(15,1.9){\line(0,1){2.1}}
\put(14.85,1.8){$\bullet$}

\put(15,4){\line(2,-3){2}}

\put(16,2.7){\tiny $-1$}

\put(13.4,1.6){\tiny $\alpha_1$}
\put(14.9,1.1){\tiny $\alpha_3\! +\! 1$}
\put(14.1,3){\tiny $\alpha_2$}
\put(14.7,4.3){$0$}
\put(12.8,0.5){$x$}
\put(17,0.5){$y$}

\put(18,2.2){$) \}$}

\put(24,1.2){\bf Fig. 1}

\end{picture}

\noindent
With the help of this rule we obtain the reconstruction of graphs in which
the number of integrations (boldface vertices) does not decrease.
However, this rule is extremely useful, since the corresponding reconstruction
of the graphs leads to variations of the
indices on the lines, which further permits one to apply
previous relations {\bf a,b} and
decrease the number of integrations.

The operator version of the integration by parts rule
(Fig. 1) has the form
\be
\lb{intp}
(2 \gamma \! - \! \alpha \! - \! \beta ) \,
\hat{p}^{2 \alpha} \hat{q}^{2 \gamma} \hat{p}^{2 \beta} =
\frac{[\hat{q}^2 , \, \hat{p}^{2 (\alpha+1)} ]}{4(\alpha +1)}  \, \hat{q}^{2 \gamma}
\, \hat{p}^{2 \beta}
\! - \! \hat{p}^{2 \alpha} \hat{q}^{2 \gamma}
\frac{[\hat{q}^2, \, \hat{p}^{2 (\beta+1)} ]}{4(\beta +1)}
\ee
where $\alpha = -\alpha_1'$, $\gamma = - \alpha_2$ and $\beta = -\alpha'_3$.
Identity (\ref{intp}) can be directly proved by using the relations for the Heisenberg algebra:
\be
\lb{intp2}
\begin{array}{c}
 [ \hat{q}^2 , \, \hat{p}^{2 (\alpha+1)} ] =
4 \, (\alpha +1) \, (H + \alpha) \, \hat{p}^{2 \alpha} \; , \\ [0.2cm]
H \, \hat{q}^{2 \alpha} = \hat{q}^{2 \alpha}  \, (H + 2 \alpha)  \; , \;\;\;
H \, \hat{p}^{2 \alpha}  = \hat{p}^{2 \alpha} \, (H - 2 \alpha)  \; ,
\end{array}
\ee
where $H := \frac{\rm i}{2}(\hat{p}_i \hat{q}_i + \hat{q}_i \hat{p}_i)$
is the dilatation operator. It follows from (\ref{intp2}) that the operators
$\{ \hat{q}^2, \hat{p}^2, H \}$ generate the algebra $sl(2)$:
\be
\lb{sl2}
[ \hat{q}^2 , \, \hat{p}^2 ] = 4 \, H \; , \;\;\;
[H , \, \hat{q}^2] = 2 \, \hat{q}^2  \; , \;\;\;
H \, \hat{p}^2 = -2 \, \hat{p}^2  \; .
\ee

\subsection{Group ${\cal H}$ and characters on ${\cal H}$.}

Note that all operator identities {\bf a,b,c} (\ref{convol1}), (\ref{zvtr2}), (\ref{intp})
are relations on the operators $\hat{p}^{2\alpha}$, $\hat{q}^{2\beta}$
and their products. So it is natural
to introduce a group ${\cal H}$ which is generated by the operators
$\{ \hat{p}^{2\alpha},\hat{q}^{2\beta}\}$ $(\forall \alpha,\beta \in {\bf C})$.
The dilatation operator $H$ belongs to the group algebra
of ${\cal H}$ in view of the first relation in (\ref{sl2}).
Consider any element of the group ${\cal H}$
\be
\lb{operator}
\Psi(\alpha_i) = \hat{p}^{-2\alpha_1'} \, \hat{q}^{-2\alpha_2} \, \hat{p}^{-2\alpha_3'} \,
\hat{q}^{-2\alpha_4} \, \hat{p}^{-2\alpha_5'} \cdots \hat{q}^{-2\alpha_{2k}}
\, \hat{p}^{-2\alpha_{2k+1}'} \, .
\ee
{}For $\sum_{m=1}^k\alpha_{2m}=\sum_{m=0}^k \alpha_{2m+1}'$ and
$\forall k \in {\bf Z}_+$, such elements form a
commutative subgroup ${\cal H}_0$ in ${\cal H}$ (see Remark 1 in Sect. 3).
The element (\ref{operator}) can be interpreted as an operator version of the 3-point function:

\unitlength=6mm
\begin{picture}(25,4)

\put(0,1){$\langle x | \Psi(\alpha_i) | y \rangle \sim$}

\put(5,1){\line(1,0){12}}
\put(10.8,3.5){$0$}

\put(6.9,0.9){$\bullet$}
\put(6.8,0.5){$z_{_1}$}
\put(9,0.9){$\bullet$}
\put(8.9,0.5){$z_{_2}$}

\put(7.9,0.7){\tiny $\alpha_3$}
\put(5.8,0.7){\tiny $\alpha_1$}
\put(8,2.2){\tiny $\alpha_2$}


\put(9.8,0.7){\tiny $\alpha_5$}
\put(10.3,2.1){\tiny $\alpha_4$}

\put(11,2.1){$\cdots$}
\put(10.5,0.7){$\cdots$}

\put(9.2,1){\line(3,4){1.8}}
\put(7,1){\line(5,3){4}}
\put(12.8,1){\line(-3,4){1.8}}
\put(15,1){\line(-5,3){4}}


\put(13.3,2.1){\tiny $\alpha_{2k}$}
\put(13.3,0.7){\tiny $\alpha_{2k-1}$}
\put(15.5,0.7){\tiny $\alpha_{2k+1}$}

\put(12.7,0.9){$\bullet$}
\put(14.8,0.9){$\bullet$}
\put(14.7,0.5){$z_{_k}$}

\put(4.5,0.6){$x$}
\put(17,0.6){$y$}

\end{picture}

\noindent
Indeed, the corresponding multiple integral is obtained from the representation
\be
\lb{3to4}
\begin{array}{ccccc}
\!\!\! \langle x | \Psi(\alpha_i) | y \rangle \! = \! \langle x | \!\!
& \!\!\! \hat{p}^{-2\alpha_1'} \;\; \hat{q}^{-2\alpha_2} & \!\!\! \hat{p}^{-2\alpha_3'} \;\;
\hat{q}^{-2\alpha_4} & \!\!\! \hat{p}^{-2\alpha_5'} \cdots  \hat{q}^{-2\alpha_{2k}} &
\!\!\! \hat{p}^{-2\alpha_{2k+1}'}  | y \rangle  \\ [-0.2cm]
& \Uparrow  & \Uparrow & \Uparrow &   \\
& \!\!\! \int \!\! d^{^D} \! z_{_1} | z_{_1} \rangle \langle z_{_1} |  &
\int \! d^{^D}\!\! z_{_2} | z_{_2} \rangle \langle z_{_2} |  &
\;\; \int \!\! d^{^D} \! z_{_k} | z_{_k} \rangle \langle z_{_k} |  &
\end{array}
\ee
if we take into account eqs. (\ref{matrep}), (\ref{matrep2}).
Note that the expression
\be
\lb{propf}
\langle x | \Psi(\alpha_i) | x \rangle = \chi_D(\alpha_i) \, \frac{1}{x^{2\sum \alpha_i-kD}} \; ,
\ee
where $\chi_D(\alpha_i)$ is a coefficient function, gives the representation
for the 2-point function, or for the propagator-type FD. The advantage of the
operator approach to the evaluation of the 3-point function (\ref{3to4})
consists in that we can apply relations {\bf a,b,c} (\ref{convol1}), (\ref{zvtr2}), (\ref{intp})
to the element (\ref{operator}) directly at the operator level, i.e.
use the commutation relations in the group algebra of ${\cal H}$
instead of making the corresponding manipulations with multiple integrals.


A remarkable fact is that one can define a trace
for the elements of the group ${\cal H}$ (generated by the operators
$\{ \hat{p}^{2\alpha},\hat{q}^{2\beta}\}$).
First, recall that the dimension regularization scheme requires the identity \cite{Hooft}
\be
\lb{rgid}
\int \frac{d^{^D} x}{x^{^{2(D/2 + \alpha)}}} = 0 \;\;\; \forall \alpha \neq 0 \; .
\ee
The extension of the definition for the integral (\ref{rgid}) at the point $\alpha = 0$
was proposed in \cite{GI} and has the form
\be
\lb{gi}
\int \frac{d^{^D} x}{x^{^{2(D/2 + \alpha)}}} =
\pi \Omega_{_D} \delta(|\alpha|) \; ,
\ee
where $\Omega_{_D} = \frac{2\pi^{^{D/2}}}{\Gamma(D/2)}$ is the area of the unit
hypersphere in ${\bf R}^D$, $\alpha = |\alpha| e^{^{i \arg(\alpha)}}$
and $\delta(.)$ is the one-dimensional delta-function.
Consider the formal integral for the 2-point function (\ref{propf}):
$$
 \int \!\! \frac{d^{^D} x}{x^{2\gamma}}
\langle x | \, \Psi(\alpha_i)
| x \rangle
= \chi_D(\alpha_i)\! \int \! \frac{d^{^D} x}{x^{2(\beta+\gamma)}} =
$$
\be
\lb{trace}
= \int \!\! d^{^D} x
\langle x | \, \Psi(\alpha_i) \hat{q}^{-2\gamma} | x \rangle
 =:  {\rm Tr} \left( \Psi(\alpha_i) \hat{q}^{-2\gamma} \right)
\pi \Omega_{_D} \delta(|\beta + \gamma -D/2|) \; .
\ee
Here $\beta = \sum_{i} \alpha_i-k\frac{D}{2}$ and $\chi_D(\alpha_i)$ is a coefficient function
for propagator type diagram which now can be interpreted
as a character for the group element
$$
\hat{p}^{^{-2\alpha_1'}} \, \hat{q}^{^{-2\alpha_2}}
\!\!\! \cdots  \hat{p}^{^{-2\alpha_{2k+1}'}} \hat{q}^{-2\gamma} \in {\cal H}_0 \; ,
$$
where $\beta + \gamma =\frac{D}{2}$ (or $\sum_{m=1}^k\alpha_{2m}+\gamma = \sum_{m=0}^k \alpha_{2m+1}'$).
The cyclic property of the trace
(\ref{trace}) $Tr(AB) = Tr(BA)$ can be checked directly.
The definition of the trace (\ref{trace}) permits one to reduce the evaluation
of propagator-type FDs (and searching for their symmetries)
to the evaluation (and searching symmetries) of vacuum FDs (for details see \cite{GI}).

\section{Applications}
\setcounter{equation}0

\subsection{Ladder FDs for  $\phi^3$ theory in $D=4$; relation to conformal
quantum mechanics}

Consider dimensionally and analytically regularized massless perturbative integrals
\be
\lb{intlad}
D_L (p_0, p_{_{L+1}}, p; \alpha, \beta, \gamma) =
\left[ \prod_{k=1}^{L} \int
\frac{d^{^D} p_k}{p^{2\alpha}_k \, (p_{_k} - p)^{2\beta}} \right]
\prod_{m=0}^{L} \frac{1}{(p_{_{m+1}} - p_{_m})^{2\gamma}} \; ,
\ee
which correspond to FDs ($x_1 = p_0$, $x_2 = p_{L+1}$, $x_3 =p$)

\unitlength=5mm
\begin{picture}(25,6.5)

\put(0.5,3){\line(1,0){12}}

\put(2.4,2.8){$\bullet$}
\put(4.5,2.8){$\bullet$}
\put(3.6,3.2){\tiny $\gamma$}
\put(1.3,3.2){\tiny $\gamma$}
\put(3.5,4.2){\tiny $\beta$}
\put(4.2,2.1){\tiny $\alpha$}

\put(5.3,3.2){\tiny $\gamma$}
\put(5.8,4.1){\tiny $\beta$}
\put(5.6,2.1){\tiny $\alpha$}
\put(6.5,4.1){$\cdots$}
\put(6,3.2){$\cdots$}
\put(6.5,2.1){$\cdots$}

\put(4.7,3){\line(3,4){1.8}}
\put(2.5,3){\line(5,3){4}}
\put(8.3,3){\line(-3,4){1.8}}
\put(10.5,3){\line(-5,3){4}}

\put(4.7,3){\line(3,-4){1.8}}
\put(2.5,3){\line(5,-3){4}}
\put(8.3,3){\line(-3,-4){1.8}}
\put(10.5,3){\line(-5,-3){4}}

\put(9.2,1.8){\tiny $\alpha$}
\put(8.8,4.1){\tiny $\beta$}
\put(9,3.2){\tiny $\gamma$}
\put(11,3.2){\tiny $\gamma$}

\put(8.2,2.8){$\bullet$}
\put(10.3,2.8){$\bullet$}

\put(0.7,2.4){$x_1$}
\put(12,2.4){$x_2$}
\put(6.3,5.6){$x_3$}
\put(6.4,-0.1){$0$}

\put(13.3,2.9){\bf $=$}

\put(15,4){\vector(1,0){9}}
\put(15,4){\vector(1,0){7}}
\put(15,4){\vector(1,0){2}}
\put(15,4){\vector(1,0){0.5}}
\put(24,2){\vector(-1,0){9}}
\put(24,2){\vector(-1,0){7}}
\put(24,2){\vector(-1,0){2.5}}
\put(24,2){\vector(-1,0){0.5}}

\put(16,2){\line(0,1){2}}
\put(16,2){\vector(0,1){1.5}}
\put(18,2){\line(0,1){2}}
\put(18,2){\vector(0,1){1.5}}
\put(20.5,2){\line(0,1){2}}
\put(20.5,2){\vector(0,1){1.5}}
\put(22.5,2){\line(0,1){2}}
\put(22.5,2){\vector(0,1){1.5}}
\put(18.2,2.8){$........$}

\put(14,4.3){\tiny $x_1$-$x_3$}
\put(14.7,1.6){\tiny $x_1$}
\put(16.3,4.3){\tiny $p_1$- $p$}
\put(16.8,1.6){\tiny $p_{_1}$}
\put(20.7,4.3){\tiny $p_{_{L}}$ - $p$}
\put(21.2,1.6){\tiny $p_{_{L}}$}
\put(20.6,2.8){\tiny $p_{_{L\,L-1}}$}

\put(23.5,4.3){\tiny $x_2 \!\!\! - \! x_3$}
\put(23.8,1.6){\tiny $x_2$}
\put(22.7,2.8){\tiny $p_{_{L+1\,L}}$}

\put(15.3,2.8){\tiny $p_{_{10}}$}
\put(17,2.8){\tiny $p_{_{21}}$}

\put(16,0.1){$(p_{_{mk}} = p_m - p_k)$}

\put(26,0.1){\bf Fig.2}

\end{picture}

\noindent
Perturbative integral (\ref{intlad}) can be graphically represented in two ways
 -- as diagrams in configuration and momentum spaces, as it is shown in Fig. 2,
where $\alpha,\beta,\gamma$ are indices on the lines in the left
diagram (in configuration space) , while in the right
diagram (in momentum space) the indices
$p_i$ indicate momenta flowing over the lines.
As we have mentioned above, these diagrams are dual to each other
(boldface vertices in the left diagram correspond to the loops in the right diagram).
The operator version of the integral (\ref{intlad}) follows from the representation
for the left diagram
$$
\begin{array}{c}
D_L (x_a; \alpha, \beta, \gamma) = (a(\gamma\, '))^{-L-1} \,
\langle x_1| \hat{p}^{-2\gamma\, '} \left( \prod_{k=1}^L \,
\hat{q}^{-2\alpha} (\hat{q}-x_3)^{-2\beta}
\hat{p}^{-2\gamma\, '} \right)  |x_2 \rangle \; .
\end{array}
$$
It is convenient to consider the generating function for the integrals $D_L$ (\ref{intlad})
\be
\lb{gener}
D_{g} (x_a;\alpha , \beta, \gamma ) =
a(\gamma\, ') \, \sum_{L=0}^{\infty} g^L \, D_L (x_a;\alpha , \beta, \gamma ) =
\langle x_1 \, |  \! \left(  \! \hat{p}^{2\gamma\, '} \!\! -
\frac{g/ a(\gamma \, ')}{ \hat{q}^{2\alpha} (\hat{q}-x_3)^{2\beta}} \! \right)^{\!\! -1} \!\!\!
| \, x_2 \rangle \; .
\ee
If the indices on the lines are related by the condition
\fbox{$\alpha + \beta = 2\gamma\, '$}, then by using properties (\ref{svR}),
(\ref{svR1}) of the inversion operator
${\cal R}$ we obtain (for details see \cite{Isa03}):
\be
\lb{dgab}
D_g(x_a; \, 2\gamma\, '-\beta , \beta, \gamma ) =
\frac{1}{(x_1^2 x_2^2)^\gamma}  \langle u \, |  \, \left( \hat{p}^{2\gamma \, '}
- \frac{ g_{\gamma \, ',\beta} }{ \hat{q}^{2\beta} } \right)^{-1} \, | \, v \rangle \; ,
\ee
where $g_{\gamma,\beta} = \frac{g}{(x_3)^{2\beta}a(\gamma)}$,
$u_i =  \frac{(x_1)_i}{(x_1)^2}- \frac{(x_3)_i}{(x_3)^2}$,
$v_i = \frac{(x_2)_i}{(x_2)^2}- \frac{(x_3)_i}{(x_3)^2}$ $(i=1,\dots,D)$.
In the case when we fix the indices on the lines as $\gamma\, ' =\beta$, the generating function
$D_g$ (\ref{dgab}) is related to the conformal function
\be
\lb{dgab12}
D_g(x_a; \, \beta , \beta, D/2-\beta) =
\frac{1}{(x_1^2 x_2^2)^{(D/2-\beta)}} \langle u \, |  \, \left( \hat{p}^{2\beta}
- \frac{ g_{\beta,\beta} }{ \hat{q}^{2\beta} } \right)^{-1} \, | \, v \rangle \; ,
\ee
and then taking $\beta =1$ we obtain the Green function for $D$-dimensional
conformal mechanics
\be
\lb{dgab1}
D_g(x_a; \, 1 , 1, D/2-1 ) = \frac{1}{(x_1^2 x_2^2)^{(D/2-1)}} \langle u \, |  \, \left( \hat{p}^{2}
- \frac{ g_{1,1} }{ \hat{q}^{2} } \right)^{-1} \, | \, v \rangle =
\ee
$$
= a(1) \, \sum_{L=0}^{\infty} g^L \, D_L (x_a;1, 1, D/2-1 )\; .
$$
Thus, we have shown that with a special choice of indices on the lines
$\alpha=\beta=1$, $\gamma = \frac{D}{2}-1=1-\epsilon$ the ladder diagrams (in momentum space):

\unitlength=7mm
\begin{picture}(15,5)

\put(5,4){\vector(1,0){9}}
\put(5,4){\vector(1,0){7}}
\put(5,4){\vector(1,0){2}}
\put(5,4){\vector(1,0){0.5}}
\put(14,2){\vector(-1,0){9}}
\put(14,2){\vector(-1,0){7}}
\put(14,2){\vector(-1,0){2.5}}
\put(14,2){\vector(-1,0){0.5}}

\put(6,2){\line(0,1){2}}
\put(6,2){\vector(0,1){1.5}}
\put(8,2){\line(0,1){2}}
\put(8,2){\vector(0,1){1.5}}
\put(10.5,2){\line(0,1){2}}
\put(10.5,2){\vector(0,1){1.5}}
\put(12.5,2){\line(0,1){2}}
\put(12.5,2){\vector(0,1){1.5}}
\put(8.2,2.8){$........$}

\put(4.3,4.3){$x_1$-$x_3$}
\put(5,1.6){$x_1$}
\put(6.5,4.3){\tiny $1$}
\put(6.8,1.6){\tiny $1$}
\put(11,4.3){\tiny $1$}
\put(11.2,1.6){\tiny $1$}
\put(10.6,2.8){\tiny $1-\epsilon$}

\put(13,4.3){$x_2$ - $x_3$}
\put(13.6,1.6){$x_2$}
\put(12.7,2.8){\tiny $1-\epsilon$}

\put(5,2.8){\tiny $1-\epsilon$}
\put(7,2.8){\tiny $1-\epsilon$}

\put(17,1.2){\bf Fig. 3}

\end{picture}

\noindent
are related to the Green function for
$D$-dimensional conformal mechanics. Moreover, according to the definition of the generating
function $D_g$, the expression $D_L$ for the ladder diagram with
$L$ loops ($L$ boldface vertices for FD in the configuration space) is obtained
in the expansion of the Green function (\ref{dgab1})
over the coupling constant $g$ (coefficient in order $g^L$).

The operator method of evaluation of Green function (\ref{dgab1}) is based on the
remarkable identity \cite{Isa03}
\be
\lb{idbest}
\frac{1}{\hat{p}^2 -  g / \hat{q}^2 }  =
\sum^\infty_{L=0}
\left( -\frac{g}{4} \right)^L \,  \left[ \hat{q}^{2\alpha} \,
\frac{(H-1)}{  (H-1+ \alpha)^{L+1}} \, \frac{1}{\hat{p}^2} \,
\hat{q}^{-2\alpha}  \right]_{\alpha^L} \; ,
\ee
where we have used the notation
$[ \dots ]_{\alpha^L} = \frac{1}{L!} \,
\left( \partial_\alpha^L \,
\left[ \dots \right] \right)_{\alpha = 0}$. The leading terms in the expansion of (\ref{idbest})
over $g$ give the identities
$$
\hat{p}^{-2} \, \hat{q}^{-2} \, \hat{p}^{-2} =
- \frac{1}{4 \, (H-1)} \, [ \log(\hat{q}^2), \, \hat{p}^{-2}]  \; ,
$$
$$
\hat{p}^{-2} \, \hat{q}^{-2} \, \hat{p}^{-2} \, \hat{q}^{-2} \, \hat{p}^{-2} =
\frac{(H-1)^{-2}}{32} \, [ \log(\hat{q}^2) , \, [ \log(\hat{q}^2) , \, \hat{p}^{-2} ]] +
\frac{(H-1)^{-3}}{16} \, [ \log(\hat{q}^2), \, \hat{p}^{-2}]  \; ,
$$
which can be proved directly and
immediately lead to analytical expressions for the ladder (box) diagrams
with one and two loops ($L=1,2$). In general, taking into account the integral representation
for the rational function of $H$ (in the right-hand side of (\ref{idbest}))
$$
\frac{(H-1)}{  (H-1+ \alpha)^{L+1}} = \frac{(-1)^{L+1}}{L!} \,
\int^\infty_0 \, dt \, t^L \, e^{t\alpha}
\, \partial_t \left( e^{t \, (H-1)} \,  \right) \; ,
$$
and using obvious properties of the operator $e^{t \, H}$:
$e^{t (H + \frac{D}{2})} \, | x \rangle \, = \, | e^{-t} x  \rangle$, we
can rewrite Green function appearing in (\ref{dgab1}) in the form
\be
\lb{phiL}
\langle u| \, \frac{1}{(\hat{p}^{2} - g_{1,1}/\hat{q}^{2})} \, |v\rangle =
 \sum^\infty_{L=0}
\frac{1}{L!} \left( {g_{1,1} \over 4} \right)^L \, \Phi_L(u,v) \; ,
\ee
where
$$
\Phi_L(u,v) = - a(1) \, \int^\infty_0 \! dt \, t^L \left[ \!
\left( \! \frac{u^{2} }{v^2} \! \right)^\alpha \! e^{t\alpha} \! \right]_{\alpha^L}
\!  \partial_t \left(
\frac{ e^{-t}}{\left(u - e^{-t}v \right)^{2} } \right)^{^{({D \over 2}-1)}} \! =
$$
\be
\lb{phiL3}
= \frac{a(1)}{u^{2(D/2-1)}} \Psi_L \left( \frac{v^2}{u^2} , 2\frac{(uv)}{u^2} \right) \; .
\ee

Thus, the result for the evaluation of the $L$-loop ladder diagram (Fig. 3) is
\be
\lb{baba}
D_L(x_1,x_2,x_3;1,1,\frac{D}{2}-1) = \left(\frac{1}{L! 4^L a(1)^{L} }  \right)
\frac{x_3^{2(D/2-L-1)}}{(x_{13}^2 x_2^2)^{D/2-1} } \Psi_L(\frac{v^2}{u^2} , 2\frac{(uv)}{u^2} ) \; ,
\ee
where
$u^2 = \frac{x_{13}^2}{x_1^2 x_3^2}$, $v^2 = \frac{x_{23}^2}{x_2^2 x_3^2}$,
$(u - v)^2 = \frac{x_{12}^2}{x_1^2 x_2^2}$ and  $x_{ab} = x_a - x_b$.

{}For $D=4-2\epsilon$ the function $\Psi_L(\frac{v^2}{u^2} , 2\frac{(uv)}{u^2} )$
(\ref{phiL3}) is expanded over $\epsilon$
$$
\Psi_L \left(\frac{v^2}{u^2} , 2\frac{(uv)}{u^2} \right) =
\sum_{k=0}^{\infty} \frac{\epsilon^k}{k!} \, \Phi^{(k)}_L(z_1,z_2) \; .
$$
where $z_1 + z_2 = 2(uv)/u^2$ and $z_1z_2 = v^2/u^2$.
The coefficient functions $\Phi^{(l)}_L$ are
\be
\lb{grn7}
\Phi^{(l)}_L =
 \sum_{f=0}^{L} \, \frac{(-\ln(z_1z_2))^f \, (2L-f)}{ f! \, (L-f)!}
 \sum_{m=0}^l (-)^m \, C^m_l \,
 {\bf Z}_{m}\left(z_1,z_2;2L +l-f \right) \; ,
 \ee
 where $C^m_l = \frac{l!}{l!(l-m)!}$ is a binomial coefficient and
 $$
 {\bf Z}_{m}(z_1,z_2;k) =
\frac{\Gamma(k-m)}{(z_1 -z_2) } \; \sum_{n_0, \dots ,n_m =1}^{\infty}
\frac{(z_1^{n_0} - z_2^{n_0})}{\left( \sum_{i=0}^m n_i \right)^{k-m} }
 \left(\prod^m_{i=1} \frac{z_1^{n_i} + z_2^{n_i}}{n_i} \right) \; ,
$$
is expressed in terms of \underline{multiple
polylogarithms}
\be
\lb{mplog}
{\rm Li}_{m_0,m_1, \dots, m_r}(w_0,w_1, \dots , w_r) =
\sum_{n_0 > n_1 > \dots > n_r > 0} \: \frac{w_0^{n_0} w_1^{n_1} \cdots w_r^{n_r}}{
n_0^{m_0} n_1^{m_1} \dots n_r^{m_r} } \; .
\ee
The first coefficient
(for $D=4$ or $\epsilon =0$) has the form \cite{DU1}, \cite{Bro2}
$$
\Phi^{(0)}_L(z_1,z_2) =  \frac{1}{z_1-z_2} \! \sum_{f=0}^{L} \!
\frac{(-)^{f}  \, (2L-f)!}{ f! \, (L-f)!} \, \ln^f(z_1z_2) \!
\left[ {\rm Li}_{_{2L-f}}(z_1) - {\rm Li}_{_{2L-f}}(z_2)\right] .
$$
and is expressed via the standard polylogarithms
${\rm Li}_{m}(w) =
\sum_{n=1}^{\infty} \: \frac{w^{n}}{n^{m}}$. The following coefficient was calculated
in \cite{Isa03}: $\;\;\; \Phi^{(1)}_L(z_1,z_2) =$
$$
\!\!\! = \frac{1}{(z_1 - z_2)} \!
\sum_{n=L}^{2L} \frac{n! \, \ln^{^{2L-n}}\! (z_1z_2) \left[ (n {\rm Li}_{_{n+1}}(z_1) -
{\rm Li}_{_{n,1}}(z_1,1) - {\rm Li}_{_{n,1}}(z_1,\frac{z_2}{z_1})) -
(z_1 \leftrightarrow z_2)  \right]}{ (-1)^n \, (2L-n)! \, (n-L)! } \, ,
$$
where the functions ${\rm Li}_{_{n,1}}(w_0,w_1)$ have been defined in (\ref{mplog}).

\vspace{0.2cm}
\noindent
{\bf Remark.}
Using conformal and scaling properties of the Green function
(\ref{dgab}) one can deduce the representation (cf. (\ref{phiL3})):
\be
\lb{cgfu}
\langle u  | \left( \hat{p}^{2\gamma} -
\frac{g(u^2v^2)^{\frac{\beta-\gamma}{2}}}{\hat{q}^{2\beta}} \right)^{-1} | v \rangle =
\frac{1}{u^{2(D/2-\gamma)}} \; \Psi^{\! (\gamma,\beta)} \!
\left( \frac{v^2}{u^2} , 2\frac{(uv)}{u^2} \right) \; ,
\ee
where the conformal symmetry requires
$\Psi^{\! (\gamma,\beta)} \! (u_1,u_2) = \Psi^{\! (\gamma,2\gamma-\beta)} \! (u_1,u_2)$. From
the representation (\ref{cgfu}) we obtain the identity
\be
\lb{ifg}
\begin{array}{c}
u^{2(D/2-\gamma)} \;
\langle u  |  \left( \hat{p}^{2\gamma} - g
\frac{(u^2v^2)^{\frac{\beta-\gamma}{2}}}{\hat{q}^{2\beta}} \right)^{-1}
| v \rangle = \\ [0.2cm]
= (u')^{2(D/2-\gamma)} \; \langle u' |
\left( \hat{p}^{2\gamma} - g
\frac{({u'}^2 \, {v'}^2)^{\frac{\beta-\gamma}{2}}}{\hat{q}^{2\beta}} \right)^{-1} | v' \rangle \; ,
\end{array}
\ee
where $u = {1 \over x_1} - {1 \over x_3}$, $v = {1 \over x_2} - {1 \over x_3}$,
$u' = {1 \over x_1} - {1 \over x_{12}}$, $v' = {1 \over x_{13}} - {1 \over x_{12}}$ and we
have introduced the concise notation
$({1 \over x_a})_i = {(x_a)_i \over x_a^2}$.
To prove eq. (\ref{ifg}) it is only needed to note the cross-ratio identities
$$
\frac{v^2}{u^2} = \frac{(v')^2}{(u')^2} = \frac{x_{23}^2 x_{1}^2}{x_2^2 x_{13}^2} \; , \;\;
\frac{(u-v)^2}{u^2} = \frac{(u'-v')^2}{(u')^2} =
\frac{x_{12}^2 x_{3}^2}{x_{13}^2 x_2^2}  \; .
$$
Now for both sides of eq. (\ref{ifg}) we make the inverse transformation
with respect to that used in passing from
eq. (\ref{gener}) to eq. (\ref{dgab}). As a result, we rewrite (\ref{ifg}) in the form
$$
x_3^{2(\gamma-D/2)} \,
\langle x_1 | \left( \hat{p}^{2\gamma} -
\frac{g \, x_3^{2\gamma} \, \widetilde{u}^{\frac{\beta-\gamma}{2}}}{\hat{q}^{2(2\gamma-\beta)}
(\hat{q} -x_3)^{2\beta}} \right)^{\! -1}
\!\!\! | x_2 \rangle =
$$
$$
= x_{12}^{2(\gamma-D/2)} \, \langle x_1 |
\left( \hat{p}^{2\gamma} -
\frac{g \, x_{12}^{2\gamma} \, \widetilde{v}^{\frac{\beta-\gamma}{2}}}{\hat{q}^{2(2\gamma-\beta)}
(\hat{q} -x_{12})^{2\beta}} \right)^{\! -1}
\!\!\! | x_{13} \rangle \, .
$$
where $\widetilde{u} = \frac{x_{13}^2 x_{23}^2}{x_1^2 x_2^2}$ and
$\widetilde{v} = \frac{x_{2}^2 x_{23}^2}{x_1^2 x_{13}^2}$.
Note that this identity is nothing but the relation on the generating functions
for the ladder diagrams (\ref{gener}). Expanding both the sides over the coupling constant $g$,
we obtain $D$-dimensional identities for the $L$-loop ladder
momentum diagrams in the order $g^L$:

\unitlength=6mm
\begin{picture}(25,5.5)

\put(-0.5,2.8){$x_3^{2(\gamma+ \gamma L -{D\over 2})} \,
\widetilde{u}^{\frac{L(\beta-\gamma)}{2}} \;  \times$}

\put(6.5,4){\vector(1,0){4}}
\put(6.5,4){\vector(1,0){2}}
\put(6.5,4){\vector(1,0){0.5}}

\put(10.5,2){\vector(-1,0){4}}
\put(10.5,2){\vector(-1,0){2}}
\put(8.2,2.8){$......$}

\put(7.5,2){\line(0,1){2}}
\put(7.5,2){\vector(0,1){1.5}}
\put(8,2){\line(0,1){2}}
\put(8,2){\vector(0,1){1.5}}
\put(9.5,2){\line(0,1){2}}
\put(9.5,2){\vector(0,1){1.5}}

\put(7.6,4.25){\tiny $\beta$}
\put(8.2,4.25){\tiny $\beta\dots$}
\put(7.6,1.5){\tiny $\widetilde{\beta}$}
\put(8.2,1.5){\tiny $\widetilde{\beta}\dots$}

\put(7,2.8){\tiny $\gamma'$}
\put(9.7,2.8){\tiny $\gamma'$}

\put(5.4,4.3){\tiny $x_1 -x_3$}
\put(6.1,1.6){\tiny $x_1$}

\put(10.2,4.3){\tiny $x_2-x_3$}
\put(10.2,1.6){\tiny $x_2$}

\put(12,2.8){$=$}

\end{picture}

\unitlength=6mm
\begin{picture}(25,4)

\put(9.5,2.8){$= \;\;\; x_{12}^{2(\gamma+ \gamma L -{D\over 2})} \,
\widetilde{v}^{\frac{L(\beta-\gamma)}{2}} \; \times$}

\put(17.5,4){\vector(1,0){4.5}}
\put(17.5,4){\vector(1,0){0.5}}
\put(17.5,4){\vector(1,0){2.5}}

\put(22,2){\vector(-1,0){4.5}}

\put(22,2){\vector(-1,0){2.5}}
\put(22,2){\vector(-1,0){0.5}}
\put(19.2,2.8){$......$}

\put(18.5,2){\line(0,1){2}}
\put(18.5,2){\vector(0,1){1.5}}
\put(19,2){\line(0,1){2}}
\put(19,2){\vector(0,1){1.5}}
\put(20.5,2){\line(0,1){2}}
\put(20.5,2){\vector(0,1){1.5}}

\put(18.6,4.25){\tiny $\beta$}
\put(19.2,4.25){\tiny $\beta\dots$}
\put(18.6,1.5){\tiny $\widetilde{\beta}$}
\put(19.2,1.5){\tiny $\widetilde{\beta}\dots$}
\put(18,2.8){\tiny $\gamma'$}
\put(20.7,2.8){\tiny $\gamma'$}

\put(17,4.3){\tiny $x_2$}
\put(17,1.6){\tiny $x_1$}

\put(21.8,4.3){\tiny $x_2-x_3$}
\put(21.8,1.6){\tiny $x_1-x_3$}


\end{picture}

\noindent
where $\beta$, $\widetilde{\beta}=2\gamma-\beta$ and
$\gamma\, '=D/2-\gamma$ are special indices on the lines and
$x_1,x_2,x_3$ parametrize external momenta.
These identities, in the special case $D=4$ ($\epsilon=0$) and
$\beta=\gamma\, '=\widetilde{\beta}=1$,
were obtained in \cite{Sokatch} and used there for
deriving many remarkable relations for various planar FDs.

\subsection{Application to Lipatov chain model}

It was shown in \cite{Lip} that a wave function for bound states
of gluons at high energies has the property of the holomorphic factorization.
The Lipatov Hamiltonian for each of these two holomorphic subsystems has the form:
$H = \sum_{i=1}^n H_{i i+1}$, where
\be
\lb{lipHam}
H_{ik} = \hat{p}_i \ln(\rho_{ik}) \hat{p}_i^{-1} + \hat{p}_k \ln(\rho_{ik}) \hat{p}_k^{-1} +
\ln (\hat{p}_i \hat{p}_k) + 2 \gamma =
\ee
\be
\lb{lipHam1}
= 2 \, \ln(\rho_{ik}) + \rho_{ik} \, \ln (\hat{p}_i \hat{p}_k) \, \rho_{ik}^{-1} + 2 \gamma \; .
\ee
Here $\gamma = -\Gamma'(1)$ is the Euler constant,
$\rho_{ik} = q_i - q_k$, where $q_i$ are complex coordinates, and momentum operators
$\hat{p}_i = - i \frac{\partial}{\partial q_i}$ are complex derivatives.
In this subsection we show how one can easily demonstrate the equality of two
expressions (\ref{lipHam}) and (\ref{lipHam1}) for the Lipatov pair Hamiltonian
by using the operator technique discussed above.

Note that the expression (\ref{lipHam1}) (up to the constant $2 \gamma$)
appears in the expansion over $\epsilon$ of the following operator:
\be
\lb{Rlip}
R_{ik}(\epsilon):=
\rho_{ik}^{1 + \epsilon} (\hat{p}_i \hat{p}_k)^{\epsilon} \rho_{ik}^{-1 + \epsilon}
= 1 + \epsilon
\left( 2 \, \ln(\rho_{ik}) + \rho_{ik} \, \ln (\hat{p}_i \hat{p}_k) \, \rho_{ik}^{-1} \right)
+ \epsilon^2 \dots \; .
\ee
Now we use the one-dimensional analog of the operator "star-triangle" identity (\ref{zvtr2})
\be
\lb{zvtr5}
\rho_{ik}^{\alpha} \, \hat{p}_i^{\alpha + \beta} \, \rho_{ik}^{\beta} =
\hat{p}_i^{\beta} \, \rho_{ik}^{\alpha+\beta} \, \hat{p}_i^{\alpha} \; \Leftrightarrow \;
\rho_{ki}^{\alpha} \, \hat{p}_i^{\alpha + \beta} \, \rho_{ki}^{\beta} =
\hat{p}_i^{\beta} \, \rho_{ki}^{\alpha+\beta} \, \hat{p}_i^{\alpha}
\; .
\ee
Then, we have
$$
\rho_{ik}^{1 + \epsilon} \, (\hat{p}_i \hat{p}_k)^{\epsilon} \, \rho_{ik}^{-1 + \epsilon} =
\rho_{ik}^{1 + \epsilon} \, \hat{p}_i^{\epsilon} \, \rho_{ik}^{-1} \,
\rho_{ik}^{1} \, \hat{p}_k^{\epsilon} \, \rho_{ik}^{-1 + \epsilon} =
\hat{p}_i^{-1} \, \rho_{ik}^{\epsilon} \, \hat{p}_{i}^{1 + \epsilon} \,
\hat{p}_{k}^{-1 + \epsilon} \, \rho_{ik}^{\epsilon} \, \hat{p}_{k}^{1} =
$$
$$
= 1 + \epsilon \left( \hat{p}_i^{-1} \, \ln(\rho_{ik}) \, \hat{p}_{i} +
\hat{p}_{k}^{-1} \, \ln(\rho_{ik}) \, \hat{p}_{k} +
\ln (\hat{p}_{i} \hat{p}_{k}) \right) + \epsilon^2 \dots \; ,
$$
and this proves the equivalence of expressions (\ref{lipHam}) and (\ref{lipHam1}).

We stress that from the point of view of the integrability of the Lipatov model
the form of the pair Hamiltonian (\ref{lipHam}) and its relation to the operator
(\ref{Rlip}) is not accidental. Indeed, the $R$-operator (\ref{Rlip}) satisfies
the Yang-Baxter equation
$$
R_{i\, i+1}(\epsilon) \, R_{i+1 \, i+2}(\epsilon+\epsilon') \, R_{i\, i+1}(\epsilon') =
R_{i+1 \, i+2}(\epsilon') \, R_{i \, i+1}(\epsilon+\epsilon') \, R_{i+1 \, i+2}(\epsilon) \, ,
$$
which can be easily proved by using the operator "star-triangle" relation
(\ref{zvtr5}). Then the complete holomorphic Hamiltonian $H = \sum_{i=1}^n H_{i i+1}$
appears in the expansion over $\epsilon$  of the monodromy matrix (in the order $\epsilon^1$)
$$
T_{(1,2\dots ,n+1)}(\epsilon) = R_{1\, 2}(\epsilon) \, R_{2\, 3}(\epsilon)
\, R_{3\, 4}(\epsilon) \cdots R_{n\, n+1}(\epsilon) \; .
$$

Finally, we should like to note that recently the solutions of the Yang-Baxter equation for
principal series of infinite dimensional representations of $SL(N)$
have been considered in \cite{Derkach}. The results of \cite{Derkach} generalize the
discussion presented in this subsection.

\subsection{Diagrams with massive propagators}

In this subsection we consider an example of the operator approach to analytical evaluation
of the 1-loop 3-point function with one massive propagator.

First, we use the automorphism
$(\hat{p}^2 \leftrightarrow \hat{q}^2$, $H \leftrightarrow -H)$ of the $sl(2)$-algebra
(\ref{sl2}) to write the first relation in (\ref{intp2}) as
\be
\lb{intp3}
[\hat{q}^{2 \beta}, \, \hat{p}^{2} ] =
4  \beta \, ( H - \beta + 1) \hat{q}^{2(\beta -1)}  \; ,
\ee
and, then, generalize it by introducing the massive parameter $m$ as follows:
\be
\lb{gr8am}
\begin{array}{c}
[(\hat{q}^2 +m^2)^{\beta}, \, \hat{p}^{2} ] =
4  \beta \, ( H + m^2 \partial_{m^2} - \beta + 1) (\hat{q}^2 +m^2)^{(\beta -1)}   \; .
\end{array}
\ee
This identity can be converted into the integral form
$$
\frac{1}{\hat{p}^{2}} (\hat{q}^2 +m^2)^{(\beta -1)}\frac{1}{\hat{p}^{2}} =
\frac{1}{4  \beta} \int_{0}^{\infty} dt e^{t( H - \beta - 1 + m^2 \partial_{m^2} )}
[\frac{1}{\hat{p}^{2}}, \, (\hat{q}^2 +m^2)^{\beta} ]
$$
from which the representation for the 3-point function is deduced
$$
\langle x_1 | \frac{1}{\hat{p}^{2}} (\hat{q}^2 +m^2)^{(\beta -1)}\frac{1}{\hat{p}^{2}} | x_2 \rangle
= \frac{a(1)}{4  \beta} \int_{0}^{\infty} dt \; e^{t( D/2 - 1)}
\frac{(e^{-t}x_2^2 +  m^2)^{\beta} - (e^{t} x_1^2  +  m^2)^{\beta} }{(e^t x_1  -x_2)^{2(D/2-1)}}  \; .
$$
Here the left-hand side is represented in the form of the perturbative integral and
we obtain the equality
$$
\int \frac{d^D k}{(k-x_1)^{2(D/2-1)}(k^2 + m^2)^{1-\beta}(k-x_2)^{2(D/2-1)}} =
$$
$$
= \frac{1}{4  a(1)} \int_{0}^{\infty} dt \; e^{t( D/2 - 1)}
\frac{(e^{-t}x_2^2 +  m^2)^{\beta} - (x_1^2 e^{t} +  m^2)^{\beta} }{\beta \, (x_1 e^t -x_2)^{2(D/2-1)}} \; .
$$
Finally, we consider the limit $D \to 4$, $\beta \to 0$ for this relation and deduce
the identity
$$
\int \frac{d^4 k}{(k-x_1)^{2}(k^2 + m^2)(k-x_2)^{2}} =
 \pi^2 \int_{0}^{\infty} dt \;
\frac{e^{-t}}{(x_1  - e^{-t} x_2)^{2}}
\log \left( \frac{e^{-t}x_2^2 +  m^2}{ e^{t} x_1^2 +  m^2}   \right)  \; ,
$$
which is important for physical applications and
corresponds to the evaluation of the 3-point one-loop FD
(in the momentum space) with one massive line:

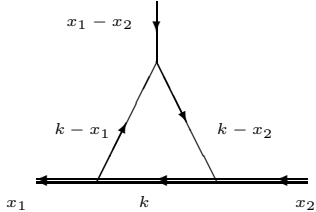
\unitlength=8mm
\begin{picture}(25,4.5)

\put(12,1.05){\vector(-1,0){4.5}}
\put(12,1){\vector(-1,0){4.5}}
\put(12,1.05){\vector(-1,0){2.5}}
\put(12,1){\vector(-1,0){2.5}}
\put(12,1.05){\vector(-1,0){0.5}}
\put(12,1){\vector(-1,0){0.5}}

\put(8.5,1){\line(1,2){1}}
\put(8.5,1){\vector(1,2){0.5}}
\put(10.5,1){\line(-1,2){1}}
\put(9.5,3){\vector(1,-2){0.5}}

\put(9.5,4){\vector(0,-1){0.5}}
\put(9.5,4){\line(0,-1){1}}

\put(8,3.6){\tiny $x_1-x_2$}
\put(9.2,0.6){\tiny $k$}
\put(7.8,1.8){\tiny $k-x_1$}
\put(10.5,1.8){\tiny $k-x_2$}

\put(7,0.6){\tiny $x_1$}

\put(11.8,0.6){\tiny $x_2$}

\end{picture}

\section*{Conclusion}

A large enough class of perturbative integrals
(which are graphically represented as FDs) generates a commutative
algebra of functions ${\cal M}$. For some subset
${\cal M}_0 \subset {\cal M}$ of generators of this algebra, the
explicit analytical expressions in terms of special functions
(of the type of multiple polylogarithms) are known. The problem of analytical evaluation
of a specific perturbative integral $I \in {\cal M}$ is now reduced to searching
for the representation of $I$ (e.g., by means of transformations
{\bf (a,b,c)} from Section 3) in terms of the elements of ${\cal M}_0$. Unfortunately,
one cannot always succeed in finding such a representation, since the subset ${\cal M}_0$
which is known nowadays,
as well as the collection of transformations (of the type {\bf a,b,c}),
do not give the possibility to speak about ${\cal M}_0$ as a complete system in ${\cal M}$.
In this case, by using the transformations {\bf (a,b,c)},
one can obtain certain relations between the elements $I_\alpha \in {\cal M}$,
which are considered as a set of functional equations for $I_\alpha$.
Thus, the problem of analytical evaluation is reduced to the extraction of
an independent set of equations and then to searching of their solution.

Now let us make a few remarks about the results presented above.

  \begin{itemize}
  \item[1.] It should be noted that the coefficient functions $\Phi_L(u,v)$ (\ref{phiL})
appear in the calculations of the 4-point functions in
the $N=4$ supersymmetric Yang-Mills theory \cite{Sok}.
  \item[2.] The proposed operator relations (\ref{zvtr2}), (\ref{zvtr5})
 not only clarify the structure of
the integrable Lipatov model (see subsection 4.2),
but also help to investigate certain generalizations
of this model (see \cite{Derkach}).
  \item[3.]
The important problem is the search for generalizations of the described algebraic formalism
in the cases of supersymmetric quantum mechanics and for massive
propagators. In the last case we have succeeded in calculating the special 3-point FD
with one massive propagator (see subsection 4.3). However, this calculation is particular. From
this point of view it would be important to calculate the coefficients $\Phi_L(u,v;m^2)$
in the expansion over $g$ of the spectral Green function for conformal mechanics:
    $$
   \langle u| \, \frac{1}{(\hat{p}^{2} - g/\hat{q}^{2}+m^2)} \, |v\rangle =
 \sum^\infty_{L=0} g^L \, \Phi_L(u,v;m^2) \; .
    $$

  \end{itemize}

\vspace{0.3cm}

I am grateful to G.~Arutyunov, S.E.~Derkachov, L.N.~Lipatov, O.V.~Ogievetsky and
E.~Sokatchev for helpful discussions and comments. I would like also to thank
the Organizers of the Conference "Quarks-2006" (Repino, May 2006) where a
preliminary version of this work was reported.

\end{document}